# Graphene-based RRAM devices for neural computing

Rajalekshmi TR, Rinku Rani Das, Chithra R and Alex James

*Abstract*—Resistive random access memory (RRAM) is very well known for its potential application in in-memory and neural computing. However, they often have different types of device-to-device and cycle-to-cycle variability. This makes it harder to build highly accurate crossbar arrays. Traditional RRAM designs make use of various filament-based oxide materials for creating a channel which is sandwiched between two electrodes to form a two-terminal structure. They are often subjected to mechanical and electrical stress over repeated read-and-write cycles. The behavior of these devices often varies in practice across wafer arrays over these stress when fabricated. The use of emerging 2D materials is explored to improve electrical endurance, long retention time, high switching speed, and fewer power losses. This study provides an in-depth exploration of neuro-memristive computing and its potential applications, focusing specifically on the utilization of graphene and 2D materials in resistive random-access memory (RRAM) for neural computing. The paper presents a comprehensive analysis of the structural and design aspects of graphene-based RRAM, along with a thorough examination of commercially available RRAM models and their fabrication techniques. Furthermore, the study investigates the diverse range of applications that can benefit from graphenebased RRAM devices.

## I. INTRODUCTION

Graphene-based resistive random-access memory (RRAM) devices have gained significant attention in recent years for their potential applications in neural computing. Graphene, a two-dimensional carbon material, has exceptional electrical and mechanical properties, making it an attractive candidate for RRAM devices. RRAM is considered one of the most promising emerging non-volatile memory, a potentially universal memory device that comes under the broad category of memristive systems [1]. The advantage of RRAM is attributed to easy to fabricate two-terminal structure that can be used to create efficient crossbar arrays, high read speeds and low area overheads. The RRAMs in crossbar can emulate multiply and accumulate (MAC) computations that are universal operation essential for implementing neural computations.

RRAM is a memory based on a resistive switching mechanism where conducting filament is created and broken due to a change of external voltage [2]. The binary RRAMs operates in two states: low resistance state (LRS) and high resistance state (HRS). Various types of electrodes and metal oxides can be used for RRAM structure. Titanium, hafnium, silicon, germanium, and nickel are the most common oxide materials, whereas silicon, silver, indium, and tantalum are familiar electrode materials used in RRAM memory devices.

Digital University, Kerala, India

Unfortunately, RRAM memory devices face various limitations with the aforementioned electrode and oxide materials [3]. For accomplishing the resistive switching property, the electrode, and conducting filament can be modified with a wide variety of materials. The electrode materials used for RRAM are divided into the following five categories: (i) elementary substance electrodes, (ii) silicon based electrodes, (iii) alloy electrodes (iv) oxide electrodes and (v) nitride based electrodes [4]. Depending on the electrode material, the number of possible states in the RRAM varies [5]. As the number of states increases, the device finds application as an analog data storage device.

In RRAM, the graphene related materials have been incorporated to increase the switching speed, retention time, endurance, and power consumption to improve the performance as a non-volatile memory [6]. Graphene provides additional properties such as transparency, flexibility, enhanced heat dissipation due to high thermal conductivity of graphene, and chemical stability. Other than these properties, as a two dimensional system, graphene can provide more than two states for the memristive device in implementing synapse for neuromorphic computing. It is reported that till now more than 16 states are possible with graphene in the memristive system [7]. Building more that two stable states in RRAMs to form analog computing systems or using them for analog storage is a open problem in RRAM-based systems.

With graphene enabled RRAMs it is expected that the higher number of states can improve the storage density and improve the reliability of the device. It is reported that RRAM devices offers switching speed of less than 10 ns, power losses of about 10 pJ, lower threshold voltage of less than 1V, long retention time of greater than 10 years, high electrical endurance with more than $10^8$ voltage cycles and extended mechanical robustness of 500 bending cycles. These advantages are complemented with its ability to tolerate high temperature variations. The graphene as interface layer act as resistive switching medium where it help to minimize the power dissipation with low contact resistance. Graphene helps to optimize the surface effect such as photodesorption and chemisorption which are varied due to increase and decrease of the temperature.

This review starts with an overview of neuro-memristive computing, graphene and its synthesis techniques. Further, the RRAM, working principle, and the resistive switching mechanism are discussed. The incorporation of graphene and graphene oxide in RRAM as an electrode and middle layer are elaborated in detail. The role of graphene in RRAM, to enhance the properties such as endurance, and retention are analyzed and the enhancement in flexibility and transparency is discussed. The progress of multilevel cell storage in RRAM is reviewed in detail. Furthermore, the commercially available RRAM models and their fabrication methods, CMOS compatibility with RRAM are also discussed.

## II. Neuro-memristive computing

### A. Memristive Devices and neural dynamics

Memristive devices have been studied for their potential to create artificial neural networks that can learn and adapt in a manner similar to biological neural networks [8]. These devices can be used to build artificial synapses that can modify their strength based on the pattern of electrical signals they receive. This is similar to how biological synapses modify their strength in response to the timing and frequency of incoming electrical signals [9]. Based on this, one potential application of memristive devices in neural dynamics is in the development of neuromorphic computing systems [10]. These systems are designed to mimic the way the brain processes information, and memristive devices could provide a way to build artificial neural networks that are more efficient and flexible than traditional computing systems [11]. This section will cover the details of differemt kinds of memristive devices, working and its viability for application in neuromorphic computing systems.

Memory-resister or Memristor, is one kind of two-terminal nano-device, considered as new-generation non-volatile memory (NVM) devices. These new computing system has proposed by Chua [12], can store information by changing the resistance of a material whereas conventional memory devices program data by change of capacitance [13]. A pinched hysteresis loop is a characteristic feature of a memristor. The loop represents the behaviour of the memristor as the voltage or current applied to it is varied has shown in figure 1. Pinched hysteresis loop is a distinctive characteristic of memristors and distinguishes them from other electronic devices such as resistors, capacitors, and inductors. Pinched hysteresis loop arises due to the inherent properties of the memristor's material and structure, which allow it to exhibit memory and resistance variations based on the history of applied voltage or current. The exact shape and characteristics of the loop depend on the specific properties of the memristor, including its materials, fabrication methods, and operating conditions. The pinched hysteresis loop of a memristor has significant implications for applications in areas such as memory devices, neuromorphic computing, and analog signal processing. It enables the memristor to store information based on its resistance state and offers unique opportunities for non-volatile memory and computing architectures.

These devices offer several advantages over conventional memory technologies such as flash, DRAM (dynamic random access memory), SRAM (static random access memory), including high density, low power consumption, and fast switching speeds [16]. They are also highly scalable, making them suitable for use in a wide range of applications from mobile devices to supercomputers . The combination of metal electrodes and insulators constructs memristor configuration.

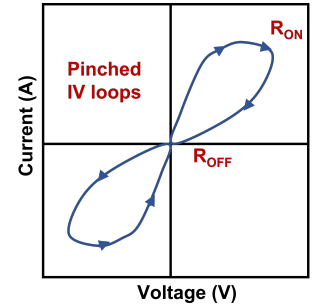

Fig. 1. An example of pinched hysterisis loop of memristor

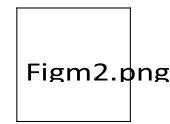

Fig. 2. (a) Memristor model according to [14] (b) traditional symbol, p-type and n-type memristors (copyright [15])

Resistive switching, phase change, spintronics and ferroelectric are the various kind of properties of memristor devices that are contributing to the development of emerging electronic technologies. Among them, resistive switching memristor(RSM) is the most common memristive device which has low power consumption, high endurance, and potential for use in neuromorphic computing [17] [18]. The applied voltage to the electrodes in RSM device creates an electric field across the metal oxide layer, causing a change in the oxidation state of the material. This oxidation state changes the resistance of the material which can be detected and used to store data. Phase change element based phase change memory (PCM) is a type of memristive device that uses a material for changing its physical state between a crystalline phase (low resistance) and an amorphous phase (high resistance) in response to heat or electric current. Spintronics memristors, or spin-torque memristors is a new type of magnetic RAM (MRAM) that works on magnetic tunnel junction (MTJ) [19] and offers high speed and high endurance performance. The resistance value has changed due to the spin of the electron and the store the data. Two ferromagnetic layers (FM) of these devices are separated by a non-magnetic (NM) layer. When an electric current is applied to the device, the spin of electrons in the magnetic layers is affected, causing a change in the resistance of the device. Ferroelectric tunnel junction (FTJ) [20] is the most significant ferroelectric memory device for neuromorphic computation, having insulating layer in between two metal electrodes. This ferroic nanostructure is comprised of an ultrathin ferroelectric barrier, and its dominant mechanism is quantum electron tunneling. In this structure, electrons are able to penetrate through the potential barrier of the ultrathin insulator. As research in this field continues to progress, memristive devices are expected to play an increasingly important role in the development of advanced computing and memory technologies.

Memristive devices are of great interest in the field of neuromorphic computing because they can be used to emulate the synaptic connections between neurons in the brain. The neural dynamics of memristive devices refers to the behavior of these devices when they are used to implement neural networks. When memristive devices are used as synapses in a neural network, their resistance values change over time in response to the input signals that they receive [21]. This behavior can be used to implement learning in the neural network, allowing it to adapt to new inputs and improve its performance over time. The dynamics of memristive devices in neural networks are highly nonlinear and can be difficult to predict [22]. However, researchers have developed models and simulations to study the behavior of these devices in neural networks.

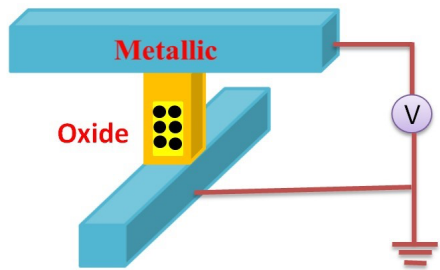

Fig. 3. Schematic of crosspoint device, showing metallic top and bottom electrodes and switching oxide. (ref [16])

*B. Memristors in crossbar*

Memristors in crossbar arrays are a type of non-volatile memory technology that hold promise for high-density, lowpower, and high-speed computing applications [23]. In a crossbar array, memristors are arranged in a grid pattern, with one set of wires running vertically and another set of wires running horizontally, forming a series of intersecting points has shown in figure 4. A single memristor crossbar refers to a specific arrangement of memristor devices in a crossbar configuration, consisting of only one layer. At each crosspoint, a memristor can be programmed to either a high or low resistance state, representing a binary 1 or 0 respectively. By applying voltage to the appropriate sets of wires, the resistance state of the memristor can be read or written. This allows for parallel access to multiple memory cells, making crossbar arrays a potential solution for memory-intensive tasks such as machine learning and artificial intelligence.

A single memristor or 1T, 1R (one-transistor, one-resistor) memristor array is typically refers to a configuration where memristors are organized in a regular grid pattern. Each memristor is positioned at the intersection of a row and a column, forming a two-dimensional array structure. The purpose of a single memristor array is to enable the simultaneous operation and interconnection of multiple memristors [24]. By arranging memristors in an array, it becomes possible to perform parallel operations, such as reading, writing, or computing, on multiple memristors simultaneously. In a 1T, 1R memristor array, each memristor is paired with a transistor. The transistor serves as the access device or switch, allowing individual memristors within the array to be addressed and read or written to [25]. The key advantage of a 1T, 1R memristor array is its high density and potential for low-power operation. By combining the storage element (memristor) and the access device (transistor) into a single unit, the overall footprint of the memory array can be reduced.

A two-memristor (2T, 2R) array consists of a configuration where two memristors are organized in a specific arrangement. There are various ways to arrange the memristors, depending on the desired application and circuit design [26]. The twomemristor crossbar array is a grid-like structure where the two memristors are positioned at the intersection of a row and a column. The rows and columns are connected to input and output nodes or other circuit elements. This configuration is commonly used in memristive crossbar arrays, where the resistance states of the memristors can be manipulated to enable or disable the connections between rows and columns [27]. Crossbar arrays are particularly relevant in applications such as memory arrays, neural networks, and digital logic circuits [28]. In a bridge memristive crossbar array, two memristors are connected in series between two nodes, forming a bridge structure. The nodes can represent inputs, outputs, or intermediate connections in a larger circuit. The bridge configuration allows for specific control over the flow of current or signal through the array. By adjusting the resistance states of the individual memristors in the bridge, it is possible to selectively enable or disable the connection between the two nodes. This can be achieved by applying appropriate voltage or current across the bridge.

Memristors in crossbar arrays also have the potential for use in neuromorphic computing, which seeks to emulate the structure and function of the human brain [23]. Memristorbased crossbar arrays can potentially perform tasks such as pattern recognition and decision-making in a highly efficient and parallelized manner. Janusz et al. developed [15]a novel neural network architecture that utilizes a compact crossbar layout of memristors, which allows us to preserve a high density of synaptic connections. Chris et al. [29] studied a memristor-based neuromorphic system for ex-situ training of multi-layer perceptron algorithms. This technique facilitates the direct translation of neural algorithm weights onto the resistive grid of a memristor crossbar. It is observed that a parallel crossbar improves the speed and power dissipation. Hu et al. [30] proposed a memristive crossbar array for high-speed image processing. It exhibits automatic memory, continuous output, and high-speed parallel computation, making it wellsuited for implementation in VLSI technology. Yifu et al. [31] developed a vertical crossbar MIM (Metal Insulator Metal) RRAM device for neuromorphic computing that is based on the 2D material ReSe$_2$. This design has been shown to exhibit improved accuracy when used in brain-inspired neuromorphic computing systems.

*C. Neuro-memristive Architectures*

The memristive circuits and computing architectures are one of the promising solutions for implementing neuromorphic computing. The memristor implementations provide various advantages such as scalability, on-chip area & power reduction, efficiency and adaptability especially for device scale-up architectures. There are existing different memristive neuromorphic architectures in the literature used for edge computing applications. The section reviews the most popular neural architectures for edge computing applications.

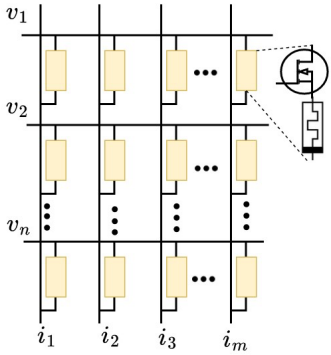

Fig. 4. ITIM Configuration for implementing DNN neural network

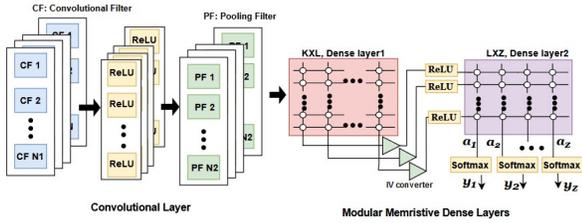

Fig. 5. CNN implementation using memristor crossbar arrays [32].

*a) Deep Neural Network (DNN):* The DNN is implemented using memristor crossbar arrays. Each DNN layer is implemented using 1T1M configuration as in Fig. 1. Each layer consists of *M* Word Lines (WLs) and *N* Bit Lines (BLs). The transistor switch enables or disables the columnwise memristor nodes. In Figure 4, $v_1$, $v_2$, ... $v_n$ from the inputs, conductance $g_{i,j}$ of memristors as weights and columns current $i_1$, $i_2$, ... $i_m$ as outputs, where *i,j* are the coordinates of the crossbar node. The output currents indicate the weighted summation of input voltages. The bias is included as an additional input line.

*b) Convolutional Neural Network (CNN):* There are several analog memristive crossbar implementations of CNN architecture [32]. Figure 5 shows the hardware implementation of CNN consisting of a convolution layer, mean pooling layer, and dense layers. The convolution filters are realized as memristive crossbars. The conductance of memristive devices is the trained weights of the convolutional filters (CF). The number of memristors in each layer is determined by the required feature maps. The features are then fed to the pooling layer circuit. The pooling layer reduces the dimensionality by performing mean-pooling operation [32]. The output of the mean-pool operation is flattened and is connected to dense layers for classification. The current-to-voltage (IV) converter block is used to convert currents to corresponding voltages. The activation functions used are ReLU and softmax.

*c) Cellular Neural Network (CeNN):* The CeNN is developed by L.O Chua and L. Yang by mimicking the features of neural network and cellular automata and finds applications in the area of image processing [33], [34]. The CeNN network in Figure 6 consists of *I×J* cells. Each cell is connected only to its neighboring cells. The connections from each cell *C(i,j)* to its neighbors is defined by cloning templates, *A(i,j;k,l)* and *B(i,j;k,l)* for feedback and feedforward connections [33], [35]. The input signal *U*, is connected to *C(i,j)* through the feedforward weights *B(i,j;k,l)*. The output of the cell $y_{k,l}$ is fed to *C(i,j)* through the feedback weights *A(i,j;k,l)*. The state equation can be mathematically expressed as [33]:

$$\frac{\mathrm{d}x_{i,j}}{\mathrm{d}t} = -x_{i,j} + \sum_{c_{k,l}} A(i,j;k,l)y_{k,l}$$
$$+ \sum_{c_{k,l}} B(i,j;k,l)u_{k,l} + I_b \quad (1)$$

where $I_b$ is the bias current, $x_{i,j}$ is the cell state and $y_{i,j}$ is the output respectively. There are various memristive implementation of CeNN in the literature [35], [36]. In figure6, the feedback and feedforward connections in CeNN network is implemented using memristor crossbar arrays.

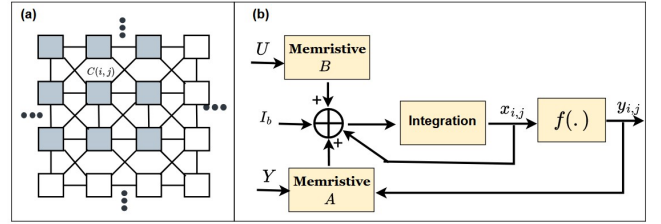

Fig. 6. (a) Structure of CeNN (b) CeNN implementation using memristor crossbar array [36].

*d) Recurrent Neural Network (RNN):* The Recurrent Neural Networks (RNNs) based methods demonstrated outstanding ability in prediction tasks using time-series data by combining large dynamical memory and adaptable computational capabilities. Long short-term memory (LSTM), the special configuration of RNN, is aimed to overcome the vanishing gradient problems in conventional RNN [37]. The memristive hardware implementation is presented in figure 7(a) [37]. The input data to the network is the concatenation of input data $x_t$, data from previous cell $h_{t-1}$ and $b_t$. The input is multiplied by a weight matrix which is the programmed conductance values of memristor crossbar array. The crossbar outputs are the input to the activation functions (either sigmoid or hyperbolic tangent) to get the gate values. $f_t$ is the output value of forget gate, $i(t)$ is the output of input/update gate, $o(t)$ denote the output from output gate and $c(t)$ denote the cell state.

The calculation time in LSTM is very heavy and timeconsuming. Echo State Network (ESN), a reservoir computing architecture, has emerged as an alternative to gradient descent training method for RNN [38]. ESN consists of an input layer where the inputs are associated with a weight matrix $w_{in}$, followed by a recurrent and sparsely connected reservoir using weight matrix $w_{res}$ and finally a readout layer associated with a weight matrix $w_{out}$. The memristive architecture of ESN reservoir layer is shown in Fig 7(b). In ESN, the output readout layer is only trained and the input and reservoir weight matrices are randomly generated and fixed throughout. The input weights are sampled from a uniform distribution $u(-a,a)$, using a scaling factor $a$ and not trained. The weights of the reservoir are sampled from $u(-1,1)$. Hence the ESN's are conceptually simple and practically easy to implement.

*e) Spiking Neural Network (SNN):* The main advantage of SNN hardware implementation is reduced power dissipation in comparison with the pulse based systems. The data signals are transmitted as spikes in Spiking Neural Networks (SNN). The SNN is based on the emulation of brain processing using particular spike events represented by Spike Timing Dependent Plasticity (STDP) . STDP is based on presynaptic and postsynaptic impulses. The implementation of SNN architectures with STDP using memristive crossbar arrays is presented in figure 8. The architecture consists of presynaptic and postsynaptic neurons connected through memristor crossbar arrays. Most cases uses Winner-Takes-All (WTA) approach for implementation [39]. Recent works introduces stochasticity by adding noise to WTA architecture. Stochasticity introduces the biological concept of probabilistic behaviour of neurons in the brain.

As discussed in the section, the field of neuromorphic computing using memristor crossbar arrays is advancing and the exploration of novel materials and devices for in-memory computing is required for improving the efficiency and scalability. The RRAM devices are promising candidates for synapses and neurons in neuromorphic circuits. The analog tunable capability of RRAM devices enable novel computing functions for the realization of neuromorphic computing. The material class for RRAM devices are from magnetic alloys, metal oxides, 2D materials and organic materials. Existing works in the literature reports that 2D material based RRAM devices have better properties compared to conventional electrode materials which enhances the characteristics of RRAM in such a way to improve its application in neural computing. The coming section reviews the mechanism of the working principle of RRAM and the use of 2D materials for enhancing the properties are discussed in detail.

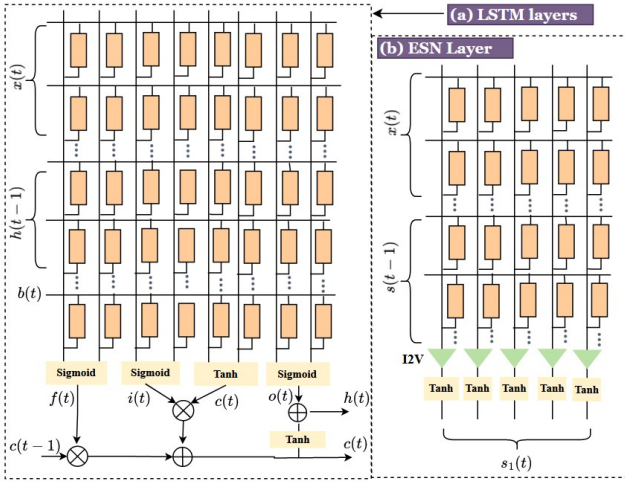

Fig. 7. (a) Memristive crossbar LSTM architecture [37], (b) ESN architecture.

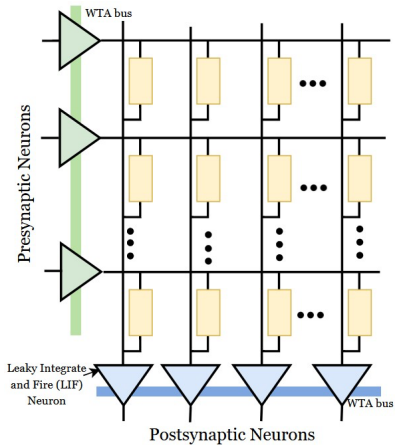

Fig. 8. Memristive Spiking Neural Network [39]

## III. Graphene and 2D materials based RRAM for neural computing

Graphene and other 2D materials have the potential to revolutionize neural computing due to their unique electrical, mechanical, and optical properties. Graphene is a single layer of carbon atoms arranged in a hexagonal lattice, and it is a highly conductive and transparent material. Other 2D materials, such as transition metal dichalcogenides (TMDs) and hexagonal boron nitride (h-BN), also exhibit interesting properties that make them promising for use in neural computing [40].

TMDs have gained significant attention in recent years due to their unique properties and potential applications in various fields, including neural computing. TMDs are a class of materials composed of transition metals (such as molybdenum or tungsten) and chalcogen elements (such as sulphur or selenium). TMDs can be used to create synaptic devices, which are fundamental building blocks of artificial neural networks [41]. TMDs exhibit excellent electrochemical properties, allowing them to function as efficient and reliable synapses. By controlling the electrical current through TMDbased synaptic devices, the strength of synaptic connections can be modulated, mimicking the synaptic plasticity observed in biological neural networks [42]. TMDs can also be utilized in the development of neuromorphic computing systems. These systems offer advantages such as parallel processing, low power consumption, and efficient data processing [43]. TMD-based devices can be integrated into neuromorphic architectures to perform tasks like pattern recognition, data analysis, and decision-making [44].

Another 2D material suitable for neural computing is the hexagonal boron nitride (h-BN) [45]. h-BN is a twodimensional material, similar to graphene, but with insulating properties. It can serve as a platform for fabricating electronic components, such as transistors, interconnects, resistive memory and sensors, with potential applications in neural computing. h-BN has been explored as a material for developing neuromorphic devices that can emulate the behaviour of biological neurons. The two-dimensional nature of h-BN allows for the integration of multiple components into compact and efficient architectures.

TABLE I
A REVIEW ON 2D MATERIALS FOR NEUROMORPHIC COMPUTING APPLICATIONS.

| Sl no. | Reference | 2D Material | Fabrication method | Target Application | switching voltage |
|---|---|---|---|---|---|
| 1 | [7] | Graphene | chemical vapor deposition (CVD) | High precision neuromorphic computing | 5.5 V |
| 2 | [46] | h-BN | CVD | Resistive memory | 0.72 V |
| 3 | [47] | $MoS_2$ | MOCVD | synapse | 0.2 V |
| 4 | [48] | $WS_2$ | RF sputtering | memristors | 1.6 V |
| 5 | [49] | $MoS_2$/Graphene | CVD | synapse | 1V |
| 6 | [50] | $MoS_2$/r-Graphene oxide | liquid exfoliation | resistive memory | 3.5 V |

Graphene-based electrodes have been shown to be biocompatible and capable of recording neural signals with high resolution and sensitivity. Additionally, graphene-based transistors have demonstrated fast switching speeds and low power consumption, making them suitable for use in neural signal processing. Another potential application is in the development of neuromorphic computing, which aims to mimic the structure and function of the human brain [7]. Graphene and other 2D materials can be used to create artificial synapses, which are the connections between neurons that allow them to communicate with each other.

Overall, graphene and other 2D materials and their combinations hold great promise for advancing the field of neural computing and could lead to the development of more efficient and powerful neural interfaces and neuromorphic computing systems. Among the 2D materials, the present review focuses mainly on the role of graphene and graphene oxide for RRAM for application in neural computing. There are still many challenges to overcome, such as improving the scalability and reproducibility of these materials and devices, before they can be widely adopted in practical applications [51]. In this section, the importance and synthesis methods of graphene are discussed in brief and a detailed analysis on the structure and working principles of RRAM is included for a better understanding of the applications of graphene basded RRAM in neural computing.

*A. Properties of graphene and the different methods for its synthesis*

Graphene is a 2D material made up of a single layer of $sp^2$ hybridized carbon atoms, arranged in a hexagonal lattice. The one atomic layer thickness makes graphene, lightweight and flexible. The strong atomic bonding with the nearest carbon atoms provides high mechanical strength to the system, greater than that of steel. Many of these properties vary based on the quality Graphene synthesized. Figure 9 shows the classification of graphene synthesis methods prevalent today. The most popular approaches include:

1) Chemical vapor deposition (CVD) - The copper or nickel substrate is heated in a reactor chamber while introducing a hydrocarbon gas (such as methane) to the chamber. These hydrocarbons react with the substrate to form graphene.
2) Epitaxial growth - Substrates similar to crystal structure of graphene (eg silicon carbide (SiC) or hexagonal boron nitride (h-BN)) can be used to grow graphene using this approach.
3) Mechanical exfoliation - The bulk crystal graphite consits for multiple layers of graphene. These layers are peeled off using tape or a sharp object.
4) Electrochemical exfoliation - The electrolyte solution is used to exfoliate graphene from graphite.

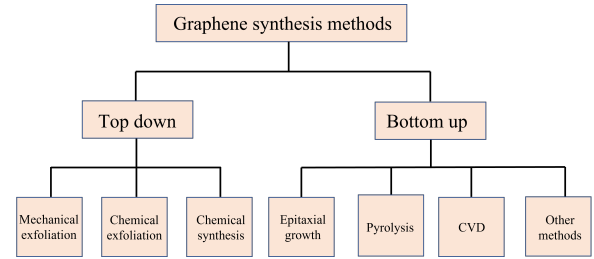

Fig. 9. Schematic representation of different methods of graphene synthesis.

5) Solvothermal sythesis - The exfoliation of graphene from bulk crystal of graphite is done in a autoclave having high pressure and temperature.
6) Thermal reduction of graphene oxide - The repeated reduction of graphene oxide by heating in hydrogen gas environment can result in graphene formation.

The discovery of graphene was through the mechanical exfoliation [52] of graphite. Different exfoliation techniques such as mechanical exfoliation, liquid-phase exfoliation [53], [54] and electrochemical exfoliation [55], [56] are used for the synthesis of graphene. In the case of mechanical exfoliation of graphene, highly ordered pyrolytic Graphite (HOPG) is used. The simplest method to exfoliate is by using a scotch tape and the graphene layer is transferred to the required substrate by sticking the tape on it. However, large-scale synthesis of graphene through this approach is time-consuming, expensive, and not practical. In practice, the use of Chemical Vapour Deposition (CVD) more commonly used to obtain high-quality graphene films [57]. In the CVD process, the gaseous reactants combine to produce the graphene layer on the substrate surface. Depending on the substrate temperature, the formation process of the sample can be controlled. With the CVD process, relatively high-quality graphene can be produced. The modern CVD techniques can be classified into LPCVD
(low-pressure CVD) and UHVCVD (Ultrahigh Vaccum CVD) (replace with PECVD, hot wall and cold wall) [58], [59].

In CVD, the deposition of a monolayer graphene on the surface of a metal substrate is relatively easy and has large area scalability potential. Several other growth techniques have been reported for graphene synthesis towards RRAM applications including atomic layer deposition (ALD) [60], solution deposition techniques [61], plasma-assisted techniques, reduction of graphene oxide [62], arc discharge [63] etc and so on. Solution coating methods such as spin coating [64], dip coating [65] and drop coating [66] offer attractive platforms for obtaining high-quality graphene films due to their low-cost and large area processability. Laser scribing technology can be used to convert GO to rGO using laser and RRAM realized using laser scribed reduced graphene oxide was reported in [67]. $CO_2$ laser-induced graphene (LIG) can be used for the fabrication of RRAM, where the graphene is transferred to polydimethylsiloxane (PDMS) from polyimide (PI) [68] and $SnO_2$ is deposited on it. This will provide a flexible RRAM device. Graphene is the thinnest material discovered to date and properties such as transparency, and flexibility make this suitable for various electronic device applications.

*B. Features and working mechanisms of Resistive Random Access Memory*

Resistive random access memory (RRAM or ReRAM) is a non-volatile memory that makes use of a material sandwiched between two metal electrode that has resistive switching properties. The resistance of the RRAM changes depending on the voltage applied across it.

The popular resistive switching material such titanium dioxide ($TiO_2$) resistance can be changed by the application of electrical current to the RRAM. The change in resistance to a high or low resistance is mapped to binary states of "0" and "1", thereby allowing digital storage. By applying voltage pulses to the RRAM electrode resistance of the $TiO_2$ film can be changed. The change in resistance is dependent on the frequency as well as the amplitude of the pulses applied. While the RRAM can be read by applying a small voltage pulse and reading the output currents without disturbing the resistance state.

Metal Insulator Metal(MIM) layer format, is used to create the structure of RRAM as shown in figure 10. The resistive switching mechanism is enabled with applications of voltage across the two terminals of RRAM to define the resistive state. The high resistance state (HRS) is considered as the OFF state, and the low resistance state (LRS) is regarded as the ON state. The switching mechanism from HRS to LRS happens through the application of external voltage. Some of the materials which exhibit this switching include the oxides of hafnium [69]–[71], titanium [72], [73], tantalum [74]–[76], zinc, nickel [77], manganese [78], magnesium [76], aluminium [79] and zirconium [80], [81]. In RRAM, the choice of electrode material is critical since it affects the switching property of the system. A small read voltage is applied to understand the system's current state (either ON or OFF) without disturbing the system's state. Since RRAM is a non-volatile memory, it will preserve the state even after removing the external voltage. RRAM can be classified into two types depending on the voltage polarity to unipolar and bipolar resistive switching. The RRAM is unipolar when the used voltage polarity is same and it is called bipolar if reverse voltage polarity is used for switching between the different resistance states (LRS and HRS).

The insulating and conducting mechanisms in the RRAM occur from the breakdown and growth of the filament on the application of an external voltage. Depending on the resistive mechanism, RRAM can be classified into (i) metal ion-based

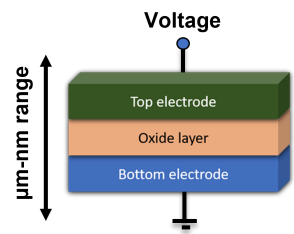

Fig. 10. Schematic structure of RRAM with Metal-Insulator-Metal layer structure

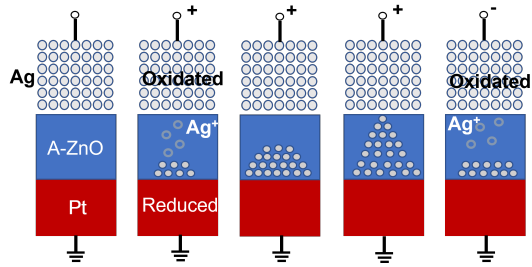

Fig. 11. Schematic of the switching mechanism of conductive bridge RRAM. (a) The pristine state of the RRAM device. (b,c) Oxidation of Ag and migration of Ag+ cations towards the cathode and their reduction. (d) Accumulating Ag atoms and Pt electrodes leads to the growth of highly conductive filaments. (e) Filament dissolution takes place by applying a voltage of opposite polarity

RRAM and (ii) oxygen vacancies filament-based RRAM. In metal ion-based RRAM, the switching mechanism happens by the migration of metal ions in the filament and the oxidation and reduction mechanism. The steps followed in the process of transitioning of conducting state to the insulating state are depicted in figure 11.

This type of mechanism happens in the case of metal electrodes such as Au, Ni or Cu at the top-level electrode. The migration of metal ions occurs through the dielectric layer, and the subsequent reduction or oxidation happens at the bottom. This will create a metal filament between the two metal electrodes through the dielectric barrier. This metal filament formation possesses the LRS state, and the disappearance of the same enables the HRS state. In figure 11, the Ag/a-ZnO/Pt RRAM cells demonstrate the resistive switching mechanism. In this case, the Ag electrode is the active element which takes part in the filament formation mechanism, and the Pt electrode is inert. The state of the RRAM devise in the absence of an external electric field is shown in figure 11(a). On applying an external voltage, the oxidation of silver takes place, and it starts to get deposited on the dielectric layer. The bottom electrode, having a negative polarity, will attract these ions and the ions get deposited on the bottom layer. The formation of metal filament through this process puts the device in the LRS state, as shown in figure 11(b)-(d). The device can be switched to the HRS state by applying the voltage in the reverse direction, as shown in figure 11(e).

In the case of oxygen vacancies-based RRAM, the resistance-switching mechanism occurs with the creation of oxygen vacancies. The reaction of oxygen ions with the anode material will create the conducting filament. The properties of RRAM will depend on the type of materials present in the top electrode, bottom electrode, and middle layer. Different substitutions of the top and bottom electrodes and middle layers with different materials can enhance the properties of RRAM. The use of 2D materials has shown an enhancement in endurance, switching speed, threshold voltage, retention time, etc. The graphene-based RRAM, shows promising results in the modification of RRAM towards better performance and for making the system a multilevel cell storage device for the application of MAC computing.

Different parameters will affect the performance of the RRAM device. This study mainly focuses on the variabilityaverse multi-level cell storage in the graphene-based RRAM system. The RRAM devices have shown a large variability due to the stochastic nature of the switching process.

## IV. GRAPHENE BASED RRAM

Improving the reliability, scalability and cost-effectiveness of the RRAM device is an important requirement for practically realizing in-memory and neural computing applications. Graphene-based RRAMs (GRRAM) have different characteristics: low power consumption, higher density, transparency, and homogeneity. GRRAM can be divided into two sub-parts: graphene RRAM and graphene oxide (GO /reduced graphene oxide (rGO)) RRAM. In graphene RRAM, graphene is used as an electrode, whereas in graphene oxide/reduced graphene oxide RRAM, GO or rGO can either be used as a dielectric layer or electrode to enhance the device's performance.

### A. Graphene as the electrode in RRAM

The main property of RRAM is the resistive switching mechanism which has various difficulties related to the selection of electrodes and the dielectric layer. The high conductivity and high surface area-to-volume ratio of graphene makes it suitable as electrodes. The power consumption is significantly less in graphene-based electrodes in RRAM compared to conventional metal electrodes in RRAM memory devices. Graphene as an electrode offers various advantages over traditional metal electrodes. The greater mechanical scalability, higher conductivity, and ultrathin nature of graphene help to design non-volatile RRAM memory devices.

Yu et al. [82] report a detailed study on resistive switching characteristics of non-volatile memory devices with nanomaterials. 2D material and nanomaterial are the extreme candidates in the nano industry where organic channels and metal electrodes decrease the transmittance value (transmittance decrease of 25 %) of the memory devices [83]. Graphene is used as electrodes, and single-wall carbon nano-tube (SW CNT) is assumed as active layers between metals in non-volatile memory devices. They implemented this memory device with ozone treatment as graphene and oxygen atoms are bonded together. The fabricated memory device revealed that it provides an acceptable transmittance value. Graphene as an electrode provides a minimum decrease of transmittance of 3.6 %, which is 11.4 % and 25 % in Au and Al. They discovered that the non-volatile memory device with graphene electrodes exhibits better conduction with high mobility of $44 cm^2V^{-1}s^{-1}$ and a switching speed of 100 ns. The graphene-based memory device performs better than metallic electrodes like Au, Al, and Ag. The graphene SWCNT memory device improves sweeping characteristics enhanced by $2 \times 10^2$ [84].

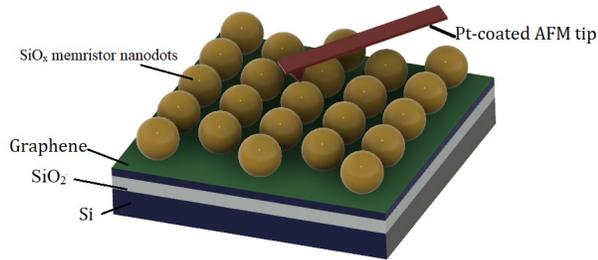

Fig. 12. Schematic structure of memristor nanostructures on metal and graphene electrodes by a block copolymer self-assembly process

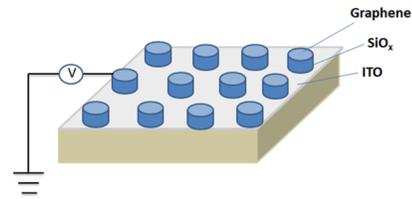

Fig. 13. Schematic diagram of graphene-SiOx-Indium Tin Oxide (ITO) device

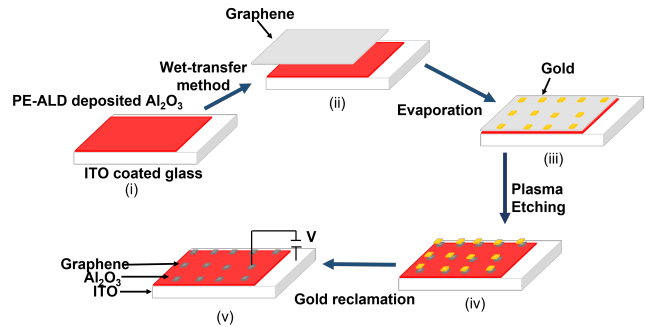

Fig. 14. Schematic structure of the Ti/$ZrO_2$/Pt RRAM device

Ji *et al.* [85] approached a design to integrate an 8 × 8 crossbar array of organic memory devices with graphene. This multi-layer graphene is an intermediate layer between insulating polyamide (IP) layers. A fabrication process integrates this device with the help of PET (polyethylene terephthalate) substrate. This device offers a high ON/OFF ratio current of $10^6$ with write-once-read-many (WORM) characteristics. The bending cycle is 10 orders larger [82] and exhibits excellent cell-to-cell uniformity. The retention time of the memory device has been controlled in the order of $10^4$. Their approach has maintained stable and reliable device characteristics without degrading the current performance. The WORM-type devices are stored the data permanently without losing any unintended data

Park *et al.* [86] demonstrated a detailed fabrication and characterization of high-density memristor nanodots with platinum and graphene electrodes by a block copolymer self-assembly process. Graphene is used as the bottom electrode, and Pt is a top electrode, where Silicon dioxide ($SiO_2$) is considered an active layer for resistive switches where the memory device has been fabricated with a minimum process cost and less complexity. The fabricated device exhibits a switching ratio of $10^2$, an endurance of 80 voltage sweeps and a unipolar switching mechanism independent on the supply voltage. The formation of a memristor on a graphene electrode is shown in figure 12.

As transparent electronics devices are in high demand for the electronics industry, Yao *et al.* [87] have configured a transparent non-volatile memory device based on $SiO_x$ active layer, indium tin oxide and graphene as bottom and top electrodes. Studies on the various device sizes are pursued to enhance the reliability of non-volatile memory. Their work revealed that the conduction filament generated in $SiO_2$ active layer maintains the constant current as the device size increases or decreases. The switching ratio ($10^5$) and electrical endurance (300 voltage sweeps) have improved compared to Park *et al.* [86]. They have also explored how the proposed device with graphene electrode offers better transparency characteristics and low retention time would be beneficial for device application. figure 13 shows the graphene-$SiO_x$-Indium Tin Oxide (ITO) device with voltage-current transfer characteristics.

A glass platform is a suitable choice for constructing transparent memory devices. The RRAM is constructed with Indium Tin Oxide as the top electrode, alumina as the functional oxide layer and graphene as the bottom electrode. The nonvolatile memory device of this composition has high transmittance of 82% in the visible region. It is stable and has non-symmetrical bipolar switching properties with low set and reset voltages (less than 1 volt). With its vertical twoterminal configuration, the device has good resistive switching performance and a strong on-off ratio (5×$10^3$) [88]. The figure representing the device structure is shown in figure 14.

A graphene-based memristive device (GMD) has been proposed by Qian *et al.* [89] and presented a comparative analysis of output performance with a Pt-based memristive device (PtmD). The Schematic Structure for Platinum-based Memristors Devices (PtMDs) and graphene-based Memristors Devices (GMDs) has shown in figure 15. The graphene electrode is integrated into $TiO_x$ by CVD fabrication method to obtain ultra-low switching power and nonlinearity. Unlike Yao *et al.* [87], they have used graphene as the bottom electrode, whereas Ti/Pt is used as the top electrode. The GMD is fabricated on poly(ethylene naphthalate) (PEN) and offers excellent retention against mechanical bending. They discovered that GMDs have less switching power compared to PtMDs, which helps to protect the device from any thermal damage. Tunable, ultralow-Power switching in memristive devices Enabled by a heterogeneous graphene–oxide Interface. The summary of RRAM devices graphene as top and bottom electrode along with typical characteristics are listed in Table II

Like Qian et al. [89], Lee et al. [90] fabricated a GS-

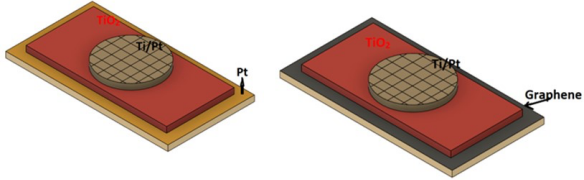

Fig. 15. Schematic Structure for (a) Platinum-based Memristors Devices (PtMDs) and (b) Graphene-based Memristors Devices (GMDs)

RRAM (GS-graphene SET electrode) memory device and compared it with a Pt-RRAM memory device. In this work, a thin monolayer graphene that serves as a SET electrode is considered to make a thin memory cell structure. The graphene SET electrode helps to store (SET) and restore (RESET) oxygen ions during the programming process. They revealed that the proposed model with a graphene edge electrode has a lower SET compliance current, low RESET current, and low programming voltages where the Pt-RRAM device cannot deal with low programming voltage or current due to degradation issues of the memory window. The efficient ion-storing capability of graphene helps reduce the power consumption 300 times more in Pt-RRAM. Metal oxide-resistive memory using graphene-edge electrodes [91] explored the performance of RRAM, where graphene is used as top and bottom electrodes. The $TiO_x$/$Al_2O_3$/$TiO_2$ dielectric layer is sandwiched between the top and bottom electrodes. The device exhibits forming-fee switching characteristics and increases the nonlinearity with a reduced value of current compliance. A stable retention value of $10^4$s, a switching ratio of $10^4$, and a greater endurance value (> 200 cycles) have been obtained for the G-I-G-based RRAM configuration.

*B. Graphene as the middle layer in RRAM*

Other than electrodes, graphene can also be used as a middle layer in GRRAM for optimizing the switching properties. The incorporation of graphene in the middle layer helps the filament growth by generating a local internal field and acts as a trapping site in the RRAM. The graphene middle layer is usually used for multilevel switching. It is reported that graphene flakes, when used as a middle layer, help trap charge and act as a storage medium.

Doh et al. [92] proposed few-layer graphene (FLG) as an active layer in field-effect devices/ferroelectric devices. They studied the effect of the graphene thickness variation to observe the electrical performance. They discovered that the device has bistable resistance characteristics with long retention values. The resistance difference ratio has decreased with the increased value of graphene film thickness. They also demonstrated that power consumption is high due to the high value of operational voltage ($V_G > 30V$). He et al. [93] proposed nanographene (NG) which acted as an active layer fabricated on a $SiO_2$ substrate. V2arious multi-level switching mechanisms have been observed, such as unipolar, bipolar, and nonpolar characteristics. Nanographene as an active layer in RRAM has several advantages, such as tunable conductivity and an easy fabrication process, unlike other materials. This research has shown a better endurance value of $10^4$ cycles, a faster-switching speed of 500 ns, and a longer retention time of $10^5$ cycles. The fabrication flow chart is shown in figure 16.

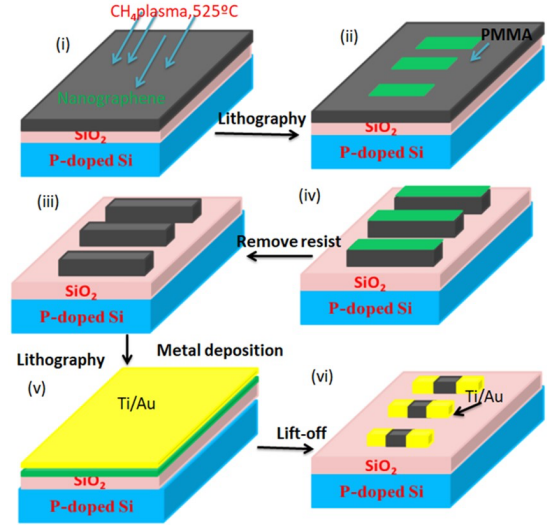

Fig. 16. Fabrication flowchart for Nanographene (NG) RRAM

Shindome et al. [94] experimented with single and multilayer graphene nanoribbon RRAM device characteristics. The drain current performance has been obtained for changing metal electrodes. They revealed that drain current is more for multi-layer graphene RRAM devices than single-layer graphene RRAM. The research also exhibits lower switching energy with a decreased value of channel width, which increases the packing density of the device. Graphene nanoribbon RRAM can possibly scale down to 30nm. Shin et al. [95] proposed the charging and discharging effect (CDE) to study the bistable switching effects in graphene devices. They also demonstrate bandgap engineering to improve the switching ratio of the device. Two different charge carriers, p-type and n-type, have been considered for this work. The proposed work revealed that the current hysteresis of p-type graphene is inverted into n-type graphene, which increases the stability of the device. The Summary of RRAM devices with graphene as an active layer along with typical characteristics are listed in Table III

## V. GRAPHENE OXIDE (GO)/REDUCED GRAPHENE OXIDE (RGO) RRAM

Graphene as a two-dimensional crystal has received more attention from researchers in the semiconductor industry due to its ultrahigh mobility, thermal conductivity, and transparency characteristics. Graphene oxide and reduced graphene oxide are the two important carbon materials mainly used in bioelectrochemical systems (BESs). Graphene oxide is a layered structure consisting of a monolayer of graphene bound to oxygen in carboxyl, hydroxyl or epoxy groups. Having a high energy band-gap of graphene oxide is possible to reduce the energy band-gap by removing the C-O bonds and offers high solubility. Graphene oxide can be deposit on any substrate due to its flexible nature. Nowadays, GO is a good insulating and semiconductor material compared to other materials and is highly used for RRAM devices. Graphene oxide-based RRAM devices have various pros compared to other materials. The RRAM device with GO can be scaled down in nano-regime and increases the packing density due to the easy fabrication process. In 2009, He *et al.* [96] first explored the RRAM device with graphene oxide (GO) thin films, which are processed by vacuum filtration method. They found that the device has a low switching voltage and offers a low $I_{ON/OFF}$ ratio, which is improved later by many researchers [97] [98]. Jeong *et al.* [99] fabricated a GO-based RRAM device prepared by the spin casting method at room temperature and found to be more reliable and flexible. This work has increased the retention and endurance of the device, which would be helpful for memory applications.

Graphene oxide (GO) can be used for nonvolatile and bistable memory devices for its high optical transparency and flexibility. Vasu *et al.* [100] studied the unipolar switching effect on reduced graphene oxide (RGO) with glass substrate to obtain a faster switching ratio and switching speed. The obtained results were better with a switching speed of $10^5$ and a high switching speed of 10 $\mu s$.

Rani *et al.* [101] implemented a cost-effective non-volatile memory behaviour in RGO memory devices for extracting better endurance and retention time. It is found that the RGO memory device exhibited an endurance value of $10^2$ and a retention value of $10^5$. Ho *et al.* [102] demonstrated a comparative analysis between RGO and GO RRAM devices for impedance spectroscopy and current-voltage analysis. The impedance spectroscopy and current-voltage analysis have been studied to determine the possible physical mechanism for resistive switching behaviour. It is observed that switching behaviour can be noticed in RGO-based RRAM devices due to its oxidation and reduction at the top electrode. The obtained results for RGO were better with the retention time of $10^6$s. However, the RGO memory device provides a large value of operating voltage of 4V, which increases the power consumption.

Pradhan *et al.* [103] proposed a nonvolatile rGO-based RRAM memory device to reduce the threshold voltage, which solves the power losses problem of the device more than Ho *et al.* [102]. Pradhan *et al.* [103] proposed an RGO RRAM device which exhibits a threshold value of less than 1V where 4V was achieved by Ho *et al.* [102]. They also checked the variability of device size, film thickness and scan time.

Kim *et al.* [104] demonstrated a transparent memory cell where reduced graphene (RGO) is placed between two ITO electrodes to observe the multi-level resistive switching purpose. This memory device offers 80% optical transmittance where the amplitude of applied pulse voltage was varied from 2V to 7V.

Lin *et al.* [105] developed a ZnO RRAM device with a capping RGO layer to study the resistive switching behaviour. They concluded that introducing the RGO layer increases the stability of the ZnO memory device with a switching ratio of $10^5$s. RGO layer act as an oxygen reservoir in ZnO memory device where ions are transit easily. On the other hand, oxygen vacancies of the RGO layer oppose reacting with Al electrodes. They also mentioned that ZnO RRAM device offers a great value of endurance of $10^8$. The Summary of RRAM devices with graphene oxide and reduced graphene oxide as an active layer are listed in Table IV

TABLE II
GRAPHENE AS TOP AND BOTTOM ELECTRODE

| Reference | Bottom electrode | Top electrode | Active layer | Substrate | Switching ratio | Endurance | Retention ratio |
|---|---|---|---|---|---|---|---|
| [106] | Graphene | Graphene | SWCNT | PET | $10^3$ | $10^2$ | $10^3$ |
| [106] | Graphene | Graphene | ZnO | Si | $10^3$ | 50 | |
| [107] | Graphene | Graphene | $TiOx/Al_2O_3/TiO_2$ | | $10^4$ | $10^2$ | $10^4$ |
| [108] | Graphene | Graphene | SiOx | Plastic | $10^6$ | $10^2$ | |
| [109] | Graphene | SiOx | Pt | Si | $10^2$ | 80 | $10^4$ |
| [110] | Graphene | Al | PMMA:P3BT | PET | $10^5$ | $10^7$ | $10^4$ |

TABLE III GRAPHENE AS AN ACTIVE LAYER

| Reference | Bottom electrode | Top electrode | Active layer | Substrate | Switching ratio | Endurance | Retention ratio |
|---|---|---|---|---|---|---|---|
| [111] | Cr/Au | Al | Graphene | $SiO_2$ | - | $10^2$ | - |
| [112] | Ti/Au | Ti/Au | Graphene | $SiO_2$ | $10^3$ | $10^4$ | $10^5$ |
| [113] | ITO | ITO | Graphene | Glass | $10^6$ | - | $10^4$ |
| [114] | Ti/Au | Ti/Au | Graphene | $Si/SiO_2$ | $10^5$ | - | - |
| [112] | Ti/Cr/au | Ti/Cr/au | Graphene nanoribbon | $Si/SiO_2$ | $10^6$ | $10^2$ | $10^3$ |

## VI. RRAM FOR MULTI-LEVEL CELL STORAGE

Multilevel cell storage in RRAM helps to increase the storage density of the memory cell without reducing the size of it. In the normal method, the cell size needs to be reduced to increase the density, which requires complex patterning techniques. In the case of multilevel cell storage, the number of bits stored per cell can be increased to n (any integer above 2), increasing the density to n times with $2^n$ number of available states in the cell. Among the different memory devices such as Spin Transfer Torque RAM (STTRAM), phase change memory etc. RRAM shows excellent scalability beyond the 10 nm technology node. The resistive switching mechanism in RRAM helps to attain different intermediate levels by varying the programming current. The size of the conducting filament in an RRAM device depends directly on the applied current. Thus, by adjusting the value of the current, different resistance states can be attained in the system.

TABLE IV
GRAPHENE OXIDE AND REDUCED GRAPHENE OXIDE AS AN ACTIVE LAYER

| Reference | Bottom electrode | Top electrode | Active layer | Substrate | Switching ratio | endurance | Retention ratio |
|---|---|---|---|---|---|---|---|
| [115] | Pt | Cu | GO | $Ti/SiO_2/Si$ | 20 | $10^2$ | $10^4$ |
| [116] | ITO | Al | GO | Glass | $10^3$ | $10^2$ | $10^9$ |
| [117] | Al | Al | GO | PET | $10^2$ | $10^2$ | $10^5$ |
| [118] | ITO | Al | GO | Glass |  | $10^3$ | $10^2$ |
| [119] | Pt | Pt | GO | $SiO_2/Si$ | $10^4$ | $10^2$ | $10^5$ |
| [120] | GO | GO | GO | PET | $10^2$ | $10^3$ | $10^3$ |
| [121] | pt | Al | Go | si | $10^4$ | $10^2$ | $10^3$ |
| [122] | Ag | Ag | Go | $SiO_2$ | 10 | - | $10^3$ |
| [118] | ITO | Al | Go | PET | $10^2$ | $10^2$ | $10^4$ |
| [103] | Al | Al | Go | glass | $10^2$ | $10^2$ | $10^4$ |
| [123] | ITO | Au | GO | Glass | 10 | $10^2$ | - |
| [124] | ITO | ITO | Go | PES | 10 | - | $10^5$ |
| [125] | Ti/Pt | Ti/Pt | go | $Si/SiO_2$ | $10^3$ | $10^4$ | $10^5$ |
| [126] | Pt | Pt | RGO | $Si/SiO_2$ | $10^5$ | - | - |
| [127] | ITO | Al/Au | Go | glass | $10^5$ | - | – |
| [128] | Ag | Au | RGO | PET | $10^4$ | $10^2$ | $10^5$ |
| [129] | ITO | ITO | RGO | Glass | $10^3$ | $10^5$ | $10^7$ |

The multilevel cell storage can be attained via different methods such as (i) Varying compliance current, (ii) Adjusting reset voltage and (iii) changing the pulse width of program/erase operation [130]. The most common method among these is the controlling of compliance current to obtain multi level cell storage. The effect of compliance current on the switching mechanism of the $Ti/ZrO_2/Pt$ is studied by Lei *et al.* [131] and the device structure is as shown in figure 8 (check). In the $Ti/ZrO_2/Pt$ device architecture, the multilevel cell storage is achieved by controlling the magnitude of the compliance current. The observed multilevel cell storage is explained using the voltage divider rule in a series circuit model. By varying the compliance current, the number of traps in the device is controlled; hence, the conductance is varied. A low voltage four-level cell storage is attained in $Ta_2O_5/TiO_2$ system by controlling the $R_L$, and $R_S$ state of the device [132]. They found that multilevel cell storage can be achieved by varying the reset voltage as well. In this work, Ru *et al.* is used as the top, and bottom electrode and the combination of $Ta_2O_5/TiO_2$ is used as the middle layer. This device achieved a 2 bit/cell storage by multi $R_H$ level operation. In another study of $HfO_2$-based RRAM system, the multilevel cell storage is achieved by controlling either $I_{set}$ or $V_{stop}$ [133].

[t!](topskip=0pt, botskip=0pt, midskip=0pt)[scale=0.3]Fig 9.png Schematic structure of the Ti/ZrO2/Pt RRAM device

In order to obtain the stable states in the multilevel cell storage system, it is important to distinguish the reference states from one another. The factors affecting the stability of resistance states are cycle-to-cycle variability, device-todevice variability, operation temperature, random telegraph noise and interstate switching variability. The study of the retention characteristics and endurance of the device will help to understand the reliability of the multiple resistance levels. It is observed that the retention time for the low resistance state highly depends on the operating current of the device [134]. With the incorporation of graphene, it is expected to obtain multiple states in the RRAM system. The property of this multi-level cell storage will enable the graphene-based systems to act as a synapse for neuromorphic computing and many other applications.

VII. COMMERCIALLY AVAILABLE RRAM MODELS AND ITS FABRICATION

For several years, researchers have demonstrated the potential of memristive devices in laboratory experiments. As a result, there have been successful demonstrations of these devices in commercial applications, with RRAM devices being particularly noteworthy in solid-state drives (SSDs) and Internet of Things (IoT) devices. Haitong *et al.* [135] proposed a memory-centric computing approach based on RRAM that leverages on-chip nonvolatile memories to perform local information processing in a highly energy-efficient manner. Three in-memory operation schemes using 3D RRAM has been developed and experimented to ensure their effectiveness and reliability, allowing for enhanced local information processing that is highly efficient and optimized for memory-centric computing systems. Wang *et al.* [136] demonstrated the integration of 1-transistor/1-resistor (1T1R) memory cells using monolayer $MoS_2$ transistors and few-layer hBN RRAMs, creating a two-level stacked 3D monolithic structure. The fabrication process was conducted at temperatures below 150 ºC. It is observed that this configuration exhibits forming-free, (¡ 1V) gradual set and reset, which is particularly advantageous for linear weight updating in neuromorphic computing. However, some renowned company has been develop various kind of RRAM devices. Adesto Technologies has recently launched a new chip family called Moneta, which utilizes CBRAM (a type of RRAM memory) technology. The Moneta family offers ultra-low power memory solutions that are designed to significantly reduce the overall energy consumption of connected devices. The chips demonstrate read and write operations at 50-100 times lower power compared to competitive solutions. The company has already begun shipping samples of the Moneta family in four different densities, including 32 Kbit, 64 Kbit, 128 Kbit, and 256 Kbit. Fujitsu recently developed ReRAM product which offers 1.5 times higher memory density compared to the existing 8 Mbit ReRAM. Other renowned foundries such as Intel, Panasonic, and Samsung have been developing RRAM technology. These companies have been investing heavily in RRAM research and development to improve the performance, reliability, and scalability of this promising memory technology.

## VIII. Graphene-based RRAM Applications

The researchers are investigating using Graphene or Graphene Oxide (GO) as electrodes or switching material of RRAM targeting in-memory computing for neuromorphic behaviour [137]–[140]. The control of resistance for multiple states by memorizing the previous state enables to mimic of biological synapses in the human brain neural network [141]–[144]. With the large development in memristive materials, an excessive amount of work is being conducted in 2D materialsbased memristors for neuromorphic computing [139], [145], [146]. The graphene crossbar variability can be used to build a unique Physical Unclonable Function (PUF), which can be used for various applications. Table V presents the review on Graphene/GO RRAM for neuromorphic Computing.

### A. Memory

The characteristic features of RRAM like simple structure, non-volatile, scalability, low power and fast operation speed makes a prominent place for future memory devices. In comparison with other materials, the 2D materials based RRAM devices offers better transparency and flexibility. The incorporation of graphene will provide more feasible and effective methods to increase the capacity of storage devices. The SET current/voltage, $I_{set}/V_{set}$, RESET current/voltage, $I_{reset}/V_{reset}$, resistance ratio $R_{OFF}/R_{ON}$, programming speed, power and retention time are the parameters for the evaluation of memory devices. Table VI shows the list of RRAM architecture in the literature with the evaluation parameters.

TABLE V
A review on Graphene RRAM for neuromorphic Computing.
*Low Resistance State $R_{LRS}$ and High Resistance State $R_{HRS}$

| Sl no. | Reference | Graphene Application | No. of Conductance states | Target Application |
|---|---|---|---|---|
| 1 | [145] [146] [141] [139], | Au/prGO/Au | 7 | ANN of size 5× 4 and 4×4 |
| 2 | | electrode of 3D vertical RRAM | 64 | XNOR |
| 3 | | RRAM fabrication with doped graphene oxide with silver | 2* | RRAM bridge synapse |
| | | Graphene Field effect Transistor | 16 | RRAM synapse |
| 4 5 6 | [142] [143] [144] | $Al_2O_3$/graphene quantum dots/$Al_2O_3$ | 2* | Synapse |
| | | Biocompatible Bilayer Graphenebased Artificial Synaptic Transistors (BLAST) | 100 | Synapse transistor |

TABLE VI
Different RRAM Architectures. MLG: Multi-layer Graphene, ITO: Indium Tin Oxide, GQD: Graphene Quantum Dots, NA: Not Available

| RRAM Structure | $I_{set}/V_{set}$ | $I_{reset}/V_{reset}$ | $R_{OFF}/R_{ON}$ ratio | SET/RESET speed | Power |
|---|---|---|---|---|---|
| MLG/$Dy_2O_3$/ITO [147] Unipolar | 1 $\mu A$/0.4V | 20 $\mu A$/0.2V | > $10^5$ | 60 ns | 4.4 $\mu W$ |
| ITO/$Al_2O_3$/Graphene [88] Bipolar | 2.1 $\mu A$/0.8V | 1.55 $mA$/-0.65V | ~3.5×$10^3$ | NA | ~1mW |
| $Al_2O_3$/GQD/$Al_2O_3$ [143] | < 5nA/1.2V | < 5nA/-1.2V | NA | NA | NA |
| ITO/GO+0.1%Ag/Al [141] Unipolar | < 4.78mA/0.8V | 2 pA/0.25V | 7.5×$10^8$ | 10$\mu s$ | NA |
| G/$SiO_x$/ITO [148] Unipolar | 2 $\mu A$/4.26V | 2 $mA$/10V | $10^4$ | 50ns | 20 mW |
| Au/prGO/Au [139] [145] Unipolar | 25 $mA$/3V | 10 $mA$/-6.5V | 10 | 10s | NA |
| TiN/$HfO_x$/Graphene [146] bipolar | < 1$\mu A$/1.27V | < 10$\mu A$/-1.37V | > 10X | 500ns | NA |

Hongbin Zhao *et al.* in [147] experimentally demonstrated that the graphene electrode layer provide high built in series resistance to exhibit good device-to-device uniformity. This exhibits narrow resistance/voltage variations in both ON and OFF states. The switching characteristics of ITO/$Al_2O_3$/Graphene RRAM is compared with ITO/$Al_2O_3$/Pt RRAM devices in [88]. The results in [88] shows that graphene shows a low SET/RESET current/voltage in comparison with conventional RRAM electrodes like Pt. A perceptron model is experimentally in [141].

Lu *et al.* [26] have developed a two-terminal memristor synapse based on a Silicon-Argon composite film. In the case of the biological synapse, the weight is varied by the release of neurotransmitters from the preneuron induced by spikes. Thus, similar to that, this memristive synapse varies its conductance by the migration of the ions upon an external electrical signal or stimuli.

### B. Neural Networks

The RRAM crossbar in-memory computing is considered to be a potential solution for implementing power efficient neural network architectures [149], [150]. The analog/digital feature of RRAM, with the ability to memorize, can be used to build artificial neural networks for neuromorphic computation [150]. Figure 17 shows crossbar architecture using RRAM devices for realizing the neuromorphic computations. The weights of neural computations are programmed onto the RRAM devices during the write mode. Only a few works have been reported in the literature using graphene/GO-based RRAM for neuromorphic computing [139], [145], [146]. Both 2D and 3D crossbar architecture with RRAM have been discussed in the literature for neuromorphic computing.

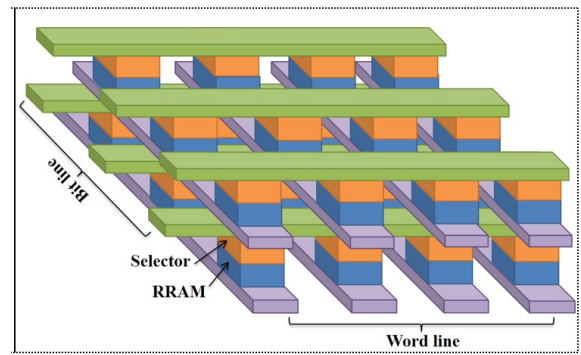

Fig. 17.    (a) 2D RRAM crossbar architecture for MAC computing (b) 3D RRAM crossbar array

Figure 17(a) shows one layer 2D RRAM crossbar schematic for neuromorphic computation. Here $n$ denotes the number of input neurons, and $m$ denotes the number of output neurons. Each column currents $i_1, i_2 ... i_m$ denote the output of MAC operation. This output voltage forms the input to the subsequent neural layers. HebaAbunahla et al. presented a novel planar analog memristor crossbar with Graphene Oxide (prGO) thin film [139], [145]. In [139], [145], the crossbar array has been fabricated and tested using the Iris dataset with an accuracy of 96.67%. 5×4 and 4×4 crossbar arrays have been fabricated, which is then used to classify the iris flower based on its petal and sepal length and width into different classes.

Batyrbek*et al.*in [146] demonstrated a 3D Vertical RRAM (VRRAM) by replacing the metal-based interconnects with graphene due to the remarkable electronic and thermal conductivities. In [146], the authors have fabricated a 416×224×8 size 3D array system. The recognition performance of the fabricated 3D Graphene RRAM (Gr-RRAM) has been tested for the MNIST dataset. The network size is 400 input, 200 hidden and 10 output neurons. The performance accuracy of Gr-RRAM is compared with platinum RRAM (Pt-RRAM), and the results show that the overall accuracy levels degrade for Pt-RRAM due to high read inaccuracy.

*C. Logic Gates*

In [146], the authors have fabricated a 416 × 224 × 8 size 3D array system. The authors demonstrated the 3D Vertical RRAM (VRRAM) by replacing the metal-based interconnects with graphene due to the remarkable electronic and thermal conductivities. An XNOR logic-inspired architecture designed to integrate 1-bit ternary precision synaptic weights into graphene-based VRRAM is introduced in [146].

*D. Cryptography*

The majority of memory based cryptographic technique for hardware security is based on Physical Unclonable Function (PUFs) [151]. The memory based PUF was first reported by G. S. Rose *et al.* in [152], [153]. The first demonstration of the memristor based PUF in hardware was presented in [154]. A large number of other memory-crossbar based PUFs have been proposed in the literature, for example metaloxide memristor based or Resistive RAM (RRAM) [153], [155]–[158] etc. The PUF methods use variations in device parameters like resistance state, switching time and threshold voltages. This unpredictable probabilistic characteristics of memristor crossbars form the basis for PUF applications. The variations in device parameter and process variations affects the current flow through the device. Any temporal or spatial variations affects all aspects of resistive switching.

RRAMs are devices capable of switching the programmed resistance from High Resistance State ($R_{OFF}$) to Low Resistance State ($R_{ON}$) in response to an applied voltage of $V_{SET}$ and vice-versa with an applied voltage of $V_{RESET}$. The variations in device parameter and process variations affects the current flow through the device. Any temporal or spatial variations affects all aspects of resistive switching [159]. The PUF can be expressed as a probabilistic procedure that maps randomized input challenges to output responses. The inherent device variations with PUFs is found to be capable of ensuring data security and device authentication for edge computing devices. A large number of RRAM-based PUFs have been proposed in the literature [155]–[157]. The variation in PUF characteristics with the properties of graphene has not been explored yet in the literature.

The stochasticity in graphene-RRAM device response has not been extensively studied in the existing literature. The SET voltage variations for cycle-cycle variations ($\sigma/\mu$) for Gr-RRAM was found to be 6.4% in [146]. The other device variations in parameters like resistance state, switching time and threshold voltages have not been considered for analysis with device-to-device and cycle-to-cycle variations. These analyses ascertain the repeatability of these devices in spatial and temporal domains for practical applications.

*E. Image Processing*

Heba *et al.* [139] presented a novel planar analog memristor crossbar with Graphene Oxide (prGO) thin film. In [139], [145], the crossbar array has been fabricated and tested using the Iris dataset with an accuracy of 96.67%. 5 × 4 and 4 × 4 crossbar arrays have been fabricated, which is then used to classify the iris flower based on its petal and sepal length and width into different classes. Batyrbek *et al.* [146] demonstrated a 3D Vertical RRAM (VRRAM) by replacing the metalbased interconnects with graphene due to the remarkable electronic and thermal conductivities. In [146], the authors have fabricated a 416 × 224 × 8 size 3D array system. The recognition performance of the fabricated 3D Graphene RRAM (Gr-RRAM) has been tested for the MNIST dataset.

IX. CMOS COMPATIBILITY

CMOS technology facing various unwanted problem due to scaling of device attributes. The semiconductor industry is planning to replace the silicon material with graphene material. Since, graphene is a conducting material and no energy bandgap present in it, it is very difficult to use graphene for digital device application due to high-off state leakage and nonsaturating drive currents. However, graphene based devices are more acceptable for low-noise amplifiers and radio-frequency (RF) in analog device application [160]. Rodriguez *et al.* [161] compared the RF behabiour between graphene based FET (GFET) and Si based MOSFET. It is observed that GFET device is more acceptable for the narrow range of drain voltage and drain current compared Si-MOSFET. Jose *et al.* [162] proposed frequency domain multiplexing of liquid-gate GFET sensor for μECoG recording purpose. The proposed work also allows hybrid integration.

Nowadays, Graphene with Si CMOS circuits can also construct together for making heterogeneous devices. The demonstration of graphene and Si CMOS hybrid circuit has reducing barriers to entry of graphene in electronics. Le Huang *et al.* [163] constructed a low temperature hybrid integrated circuit where graphene devices and Si-CMOS circuit integrated together. Carlo *et al.* [164] designed relaxation oscillators using a Graphene Field Effect Transistor (GFET), Si CMOS D latch and timing RC circuit for. It is observed that the introduction of graphene material in Si-CMOS logic circuit has improved the circuit complexity and also added the other device functionality. Zhiping *et al.* [165] proposed CMOS compatible allmetal-nitride RRAM based on Aluminium nitride (AIN). It is observed that the proposed device provides lower operation current of 100 μA, retention value $3 \times 10^5$ and endurance value of $10^5$ Hz. Chih *et al.* [166] proposed a cost competitive OneTransistor-N-RRAM (1TNR) array architecture for advanced CMOS technology where one committed transistor controls the access of one RRAM. It is observed that 1TNR array architecture provide less leakage current than the cross-point array. So, there is possibility that graphene based RRAM memory devices can be considered in CMOS technology soon.

## X. CHALLENGES AND FUTURE SCOPE

Due to its unique and interesting features, graphene has surpassed all other nanomaterials in terms of its use in electronic devices. Additionally, it was shown that graphene's greater mobility, less light absorption, and excellent mechanical qualities enhance the functionality of transparent flexible electronic devices. The difficulty is that the cost of manufacturing of graphene will increase the overall price of the device. The transfer of graphene from one substrate to another without causing any damage is a tedious process, which require the need of sophisticated instruments. Efficient methods need to be implemented to overcome these drawback.

The past several years have seen a substantial increase in research into new memory technologies, and numerous prototype RRAM products have been created to show the potential for high-speed and low-power applications. The CMOS compatibility and ability to fabricate in smaller dimensions make the RRAM a suitable candidate for device applications. A high endurance is reported in graphene based RRAM device. Till date, in a single RRAM device, no technology has reported with fast switching, low power, and stable operation simultaneously. In graphene based RRAM device also the properties need to be enhanced for a better performance of the device.

## XI. CONCLUSION

This review article offers an insightful look into the topic of developing memory technologies by giving a concise overview of the development of memory architecture, the current trends, and the constraints. The importance of graphene based RRAM, as well as its structure, operation, and classification, have all been highlighted in a thorough discussion. The methodology and a detailed investigation on the MLC capabilities of RRAM have been presented. It is proposed that the graphene-based RRAM can be used for multilevel cell storage. This modified memory device, with 2D material can be used as a synapse. Along with this, the implementation of graphene based RRAM for various important applications such as hardware security and neuromorphic computing have been highlighted.